\documentstyle{europhys}
\def\stars{\bigskip\centerline{***}\medskip}

\newif\ifboo \boofalse

\def\Review#1{\boofalse{\it #1},}
\def\Name#1{{\sc #1},}
\def\Vol#1{\ifboo Vol. {\bf #1}\else{\bf #1}\fi}
\def\Year#1{\ifboo #1\else(#1)\fi}

\def\Page#1{\ifboo {\rm p. #1}\else{\rm #1}\fi}
\begin{document}
\euro{}{}{}{}
\date{\today}

\title
{
Mode-locking of incommensurate phase by quantum zero point energy
in the Frenkel-Kontorova model
}

\author
{Bambi Hu,\inst{1,4} 
Baowen Li,\inst{2,1}\footnote{To whom correspondence should be addressed, 
Email: phylibw@nus.edu.sg}
and Hong Zhao\inst{3,1}
} 
\institute{
\inst{1}Department of Physics and Centre for Nonlinear Studies,
 Hong Kong Baptist University, China \\
\inst{2} Department of Physics, National University of Singapore, 119620
Singapore\\
\inst{3} Department of Physics, Lanzhou University, 730000 Lanzhou, China\\
\inst{4} Department of Physics, University of Houston, Houston TX 77204
USA\\
}

\rec{ }{ }
\pacs{
\Pacs{64}{70Rh}{Commensurate-incommensurate transitions}
\Pacs{63}{70+h}{Statistical mechanics of lattice vibrations and
displacive phase}
\Pacs{05}{50+q}{Lattice theory and statistics}
}

\maketitle

\begin{abstract} 
In this paper, it is shown that a configuration modulated system described 
by the
Frenkel-Kontorova model can be locked at an incommensurate phase 
when the
quantum zero point energy is taken into account. It is also found that the
specific heat for an incommensurate phase shows different
parameter-dependence in sliding phase and pinning phase. These findings
provide a possible way for experimentalists to verify the phase transition
by
breaking of analyticity.

\end{abstract}

The Frenkel-Kontorova (FK) model was proposed to study 
successive commensurate-incommensurate phase transitions observed in many configuration
modulated systems such as low-dimensional conductors having
charge density waves  and some ferroelectric materials (see recent 
review \cite{Braun1}). It has also been used widely to study transmission in Josphson
junction and atomic-scale friction -nanoscale tribology
\cite{friction,Braun2}. More recently, this model has been employed to
study transport properties of vortices in easy flow
channels\cite{Besseling} and strain-mediated interaction of vacancy lines
in a pseudomorphic adsorbate system\cite{Erwin}.
 
The FK model consists of a chain of particles having 
nearest-neighbor harmonic interaction and adsorbed in an external periodic
potential. The Hamiltonian of the FK model is

\begin{equation}
\label{1}{\cal H}=\sum_i[\frac{p_i^2}{2m}+\frac \gamma
2(x_{i+1}-x_i-a)^2-\frac V{(2\pi )^2}\cos\frac{2\pi x_i}{b}], 
\end{equation}
which can be cast into a dimensionless one 
\begin{equation}
\label{2}H=\sum_i[\frac{P_i^2}2+\frac 12(X_{i+1}-X_i-\mu )^2-\frac k{(2\pi
)^2}\cos 2\pi X_i], 
\end{equation}
where $P_i$ and $X_i$ are the dimensionless classical momentum and
position of the $i$th particle, $k=\frac V{b^2\gamma }$ is a
dimensionless coupling constant and $\mu $ the
dimensionless lattice constant at vanishing $k$. The most important 
quantity of interest is the ratio ($\rho$) of the mean distance between
neighboring particles
and the period of the external periodic potential (one)
\begin{equation}
\label{3}
\rho =\lim _{N\rightarrow \infty }\frac{X_N-X_0}N. 
\end{equation}
Because the competition between two length scales the  FK model exhibits abundant interesting phenomena\cite{Braun1,Aubry,Peyrard}. Among many others is the  mode-locking which manifests itself in the
`devil's staircase' \cite{Aubry}, i.e.,
the dependence of $\rho $ on $\mu $ is described by a highly pathological
function. Namely, $\rho(\mu)$ has the property that for all rational
number $\nu$, there exist real numbers $\mu (\nu _-)$ and $\mu (\nu
_+)$ such that $\rho (\mu )=\nu $ if $\mu (\nu _-)<\mu <\mu (\nu_+).$
These mode-locking intervals show up as horizontal plateaus in $
\rho (\mu )$. The widths of the plateaus are determined by the
Farey sequence, i.e. the widest plateau between two
plateaus at $\rho =p/q$ and $\rho =p^{\prime }/q^{\prime }$ is the plateau
at $\rho =(p+p^{\prime })/(q+q^{\prime }),$ where $p/q$ and $
p^{\prime }/q^{\prime }$ are rational fractions. Accordingly, one 
can construct the Farey tree of the rationales
with successive order starting from $0/1$ and $1/1$. Therefore, assuming
$k$  is a temperature- and pressure-dependent parameter, 
the model will exhibit a lock-in commensurate phase at relatively high order by changing temperature and/or pressure. This mechanism has been used to explain experimental results  
observed in thiourea \cite{thiourea} and in epitaxial thin film \cite{film}. 

However, to obtain a reasonable and more accurate result one should consider quantum 
effect, in particular in low temperature regime. This problem becomes more and more important 
when people starting working on nano systems at very low temperature such as nano-tribology.
One may ask: What happens to the commensurate and incommensurate 
mode-locking when the quantum effect is taken into account. In other words,
can devil's staircase survive the quantum fluctuations? 
This is not clear up to now. In fact, we have only  
limited knowledge coming from the continuum model, i.e. the 
sine-Gordon model,  which is valid only for very small $k$.
In the continuum limit,
Bak and Fukuyama\cite{Bak}  discussed the local stabilities of the 
commensurate phases. They found that
quantum effect would destroy the normal mode-locking staircase
if the quantum fluctuation is 
large enough. However, for a more general case of $k$ the picture is 
still incomplete.

On the other hand, the study of Borgonovi {\it et al}\cite{Borg89} by
using the quantum Monte Carlo method shown that the quantum fluctuations,
similar to the thermal fluctuations\cite{Vallet}, smears out the
discontinuity of hull function in classical FK model.  One of the
significant
results is that the quantum effects, mainly from tunneling,  
renormalize the standard map to an
effective sawtooth map. This phenomenon was also recovered by Berman {\it et al}
\cite{Berman} 
by using the method
of mean-field theory and Hu, Li and Zhang\cite{HLZ98} by using the
squeezed state approach. Moreover, Hu and
Li\cite{HL99} found that although the quantum effects smear out the
breaking of analyticity transition, the remnant of this transition is
still discernible in the quantum FK model, which is
demonstrated by the crossover of the ground state wave function from an
extended one to a localized one as the coupling constant is increased.
The transition also shows up in other relevant parameters \cite{HL00}.

In this paper we concentrate on effect of  quantum zero point energy. The other effects such 
as  quantum tunneling will be neglected. This approximation is valid provided that $\tilde{\hbar}$ is not too large\cite{Borg89,HL99}. As we will see soon that the quantum zero point energy can lead to interesting results. For small $k$, all the commensurate 
phases seem to be destroyed. This agrees with Bak and 
Fukuyama's finding\cite{Bak}. For large
$k$ the situation becomes more complex. The plateaus can be
destroyed and enhanced depending on the system's parameters. More interestingly, for
suitable scale of the quantum effect\footnote{In the following, the ``quantum effect'' simply means the quantum zero point energy effect.} the system may be locked into
incommensurate phases. This mode-locking of incommensurate phase has a
very special meaning in laboratory experiment. 

In the low temperature regime, we assume that the quantum effect just causes the particle a small fluctuation around its equilibrium position. The
phonon spectrum is calculated by linearizing the
system around its equilibrium configuration, it is determined by
\begin{equation}
\label{4}{\bf B}-\omega ^2{\bf I}=0, 
\end{equation}
where ${\bf B}$ is a constant matrix with elements 
$$
B_{ij}=\frac{\partial ^2H}{\partial X_i\partial X_j},
$$
and ${\bf I}$ is a unit matrix. 
The ground state energy is considered as the classical ground state
energy plus the quantum zero point energy: 
\begin{equation}
\label{5}f=v_0+f_0 
\end{equation}
where

\begin{equation}
\label{6}v_0=\frac 1N\left[ \sum_{i=1}^N\frac 12(X_{i+1}^e-X_i^e-\mu
)^2-\frac k{(2\pi )^2}\cos 2\pi X_i^e\right], 
\end{equation}
is the classical ground state energy and 
\begin{equation}
\label{8}f_0=\frac 1N\sum_{l=1}^N\frac 12\omega _l 
\end{equation}
the quantum zero point energy at the ground state. Here $X_i^e$ is the equilibrium position of $i$'th particle in classical ground state. In Eq.
(5) $v_0$ is measured in unit of $b^2\gamma $ while $f_0$  in
unit of $\hbar \omega _0$, where $\omega _0^2=\gamma /m$ and $\hbar $ is
the Plank constant. Thus the ratio 
$
\tilde{\hbar }=\frac{\hbar \omega _0}{b^2\gamma } 
$
is a measure of the
quantum zero point energy.

Throughout this paper, except special indication, $600$ particles with 
periodic boundary condition were used in our
numerical calculation. Since the system is symmetric about $\mu =0.5$, it 
is sufficient to study the interval $\rho\in[0,0.5]$. To find a stable 
phase of the system at a fixed $\mu$, i.e. the phase has a minimum $f$, we compared three hundred phases of $\rho =i/600,$ $i=1,2,\cdots ,300.$
 
The classical ground state energy $v_0$ in Eq. (5), is determined by
parameters $k$, $\rho$ and $\mu$. It has minimums
at rational values of $\rho$. $\rho$ versus $\mu $  shows
the devil's staircase. The quantum
zero point energy $f_0$, however, depends only on the parameters $k$ and $\rho$. 
The dependence of $f_{0}$ on $\rho $ is shown in Fig. 1 for different values of $k$, 
where one can see that $f_{0}$ takes
maximums at rational values of $\rho$. It has
smaller value for incommensurate phases (irrational values of $\rho$). Furthermore, the positions of the
peaks in the figure follow the Farey sequence, i.e., between two
peaks at $\rho =p/q$ and $\rho =p^{\prime }/q^{\prime }$ there exists a
peak at mediant $\rho =(p+p^{\prime })/(q+q^{\prime })$ and it satisfies the condition $\min
\{f_0(p/q),f_0(p^{\prime }/q^{\prime })\}<f_0[(p+p^{\prime })/(q+q^{\prime
})]<\max \{f_0(p/q),f_0(p^{\prime }/q^{\prime })\}$. The peaks at high order
rationales are indeed flattened.  The orders of rationales above which the peaks are flattened are determined by $k$. And it is found that more and more peaks at rationales of higher orders
are flattened as $k$ is increased. The higher the order of the rational, the earlier the
peak at the corresponding position is flattened.

The stable phase of the system is a consequence of the
competition between the two terms in Eq. (5),  $v_0$
favorites a commensurate phase but $f_0$ an incommensurate phase. The final phase of the 
system depends on the parameters $\tilde{
\hbar }$, $k$, and $\mu $. In the case of $\tilde{\hbar }\ll 1$, i.e., in
the classical limit, $v_0$ gives the major contribution, and it
determines the minimum energy phases of the system. Thus the devil's
staircase mode-locking in the classical model survive the quantum effect.
 On the other hand,
in the case of relatively large 
$\tilde{\hbar }$, the quantum effect dramatically changes the
classical
mode-locking structure of the system. 

In Fig. 2, we plot $\rho$, at which $f$ has minimum, as a function of 
$\mu$ for different values of $\tilde{\hbar}$ 
at $k=1.1$. The thick line in the figure is the result of $\tilde{\hbar }
=0$. The plot shows that
those plateaus at $\tilde{\hbar }=0$ evolve in two different ways as
$\tilde{\hbar }$ is increased. Some plateaus at the rationales, whose
corresponding peaks in the $f_0-\rho $ graph is flattened, may persist and
even be enhanced in a certain interval as $\tilde{\hbar }$ is increased.
For example, in Fig. 1 no peaks show up in the interval $
0<\rho <1/6$ in curve of $k=1.1$. However, as $\tilde{\hbar}$ increases, the plateaus at $1/7,$ $1/8$, $1/9$ and $1/10$ become more and more evident as is demonstrated in Fig. 2.
Similarly, the plateaus at $\rho =2/11$ and $\rho =3/17$ expand tremendously 
as $\tilde{\hbar }$ is increased, while Fig. 1 does not show any peaks at the 
corresponding positions. 
The expanding rate of the survived 
plateaus is different. The plateaus at rationales of higher orders expand faster.
On the other hand, the plateaus, at the positions
where $f_{0}-\rho $ graph shows peaks, are flattened quickly with the
increase of $\tilde{\hbar }$. These plateaus locate at the positions
with rationales of lower order. 
The most interesting result is those new plateaus  
created at the positions of irrationals. They grow up quickly, and become
wide plateaus in the $\rho -\mu $ graph, see e.g., the plateaus designated by the
continued fraction representation of the numbers in Fig. 2. These plateaus do not exist at $\tilde{\hbar}=0$ (thick line).

According to the number theory, a number $p/q$ can be written as a continued fraction
$$
\frac pq=\frac 1{a_0+\frac 1{a_1+\frac 1{a_2...}}}\equiv[a_0a_1a_2...]. 
$$
A rational is expressed as a finite
continued fraction and an irrational a nonterminating continued fraction.
For numerical calculation we use the Fibonacci sequence to
approach the irrationals, for example we use $\rho =233/610$ to approximate the
golden mean $[21111...]$ and $\rho =144/521$ the silver mean $%
[31111...]$\cite{note}.
In these cases our computation were performed with $610$ 
and $521$
particles, respectively. We call the plateau at an irrational number as
`{\it an irrational plateau}' and refer to this kind of mode-locking as 
incommensurate phase mode-locking in the following. Surprisingly, the
irrational plateau is obviously related to the `irrationality' of the
number (in the sense that the golden mean is the most irrational `irrational' 
and then the silver mean and copper mean etc), i.e., the more `irrational' the 
number, the easier the system can be locked at. The physical meaning of this 
statement are two folds. First,
by fixing $k$ and increasing $\tilde{\hbar }$ from zero, the first
locked irrational plateau appears at the golden mean $\rho =[21111...]$ 
followed
by the silver mean $\rho =[$$31111...]$, and then the copper mean $\rho
=[4111...]$ and so on. The widths of the plateaus are ordered respectively, i.e., 
for a fixed $\tilde{\hbar },$ one has $
l([21111...])>l([31111...])>l([41111...])...,$ where $l(\rho )$ denotes the
width of the plateau at $\rho $. Second, for a fixed value of $k$, the
number of the irrational plateaus is finite and it increases as $k$ is increased.
In the case of $k>1.5$ most of the mode-locking plateaus
seem to be located at the positions of rationals of high order. The only
irrational plateau can be recognized is that one locating at the position of
the golden mean. With the decrease of $k$ the irrational plateaus at the
silver mean, the copper mean, the iron mean, $\ldots $ , and the irrational
numbers close to these well-known irrational numbers such as $[22111...],$ $
[21211...],$ $[32111...]$ and $[31211...]$ etc. appear gradually. At
$k=1.1$ we find that the plateaus at the silver mean and the copper
mean already shown up, as is shown in Fig. 2. Our calculations indicate that
at $k=0.9$ an irrational plateau also appear at the position of the iron
mean $[51111...]$. The irrational plateaus at $[61111...]$, $
[71111...],$ and $[81111...]$ show up too at $k=0.7$, but the plateaus
at rationals can not be seen any more when $k$ below this value 
at relatively large $\tilde{\hbar}$. This result, i.e.
for small $k$ all the rational plateaus are
flattened as the quantum effect is incrased,
agrees with the discovery by Bak and Fukuyama \cite{Bak} 
qualitatively.

The mode-locking of the incommensurate phase shown in Fig. 2 has a special meaning
to experimentalists.
It might allow us to detect the signature of
the phase transition by breaking of analyticity\cite{Aubry,Peyrard} in 
laboratory. The idea is
based on the measurement of the specific heat of an incommensurate phase.
The specific heat of the system is given by
\begin{equation}
C_v=\sum_{l=1}^N(\frac{\omega _l}{T})^2\frac{e^{\omega 
_l/T}}{(e^{\omega_l/T}-1)^2}, 
\label{specheat}
\end{equation}
where $T$ is the system temperature measured in unit of $\hbar \omega 
_0/k_B$, and 
$k_B$ is the Boltzmann constant. $C_v$ is a function of the parameter $\rho$,  
$k$ and $T$. It does not depend on $\mu $. In a commensurate
phase e.g. $\rho=1/3$, we find that $C_v$ decreases exponentially with $k$. 
In an incommensurate phase the specific
heat shows a quite different behavior. It  also decreases exponentially with 
$k$ for $k>k_c(=0.9716...)$. But it changes
insensitively to $k$ for $k<k_c$. This can be easily seen from 
Fig. 3. Therefore, experimentally when the system
is found to be in a locked incommensurate phase, one shall be able to observe this 
different parameter-dependence behavior in specfic heat by changing external parameters such as pressure. This might be useful to verify the existence of $k_c$ in real experiment and detect the manifestation of phase transition by breaking of analyticity.

In summary, we have studied the effect of the quantum zero point 
energy on commensurate and incommensurate phases. If it  
is negligible the FK model shows the usual devil's staircase mode-locking, and the 
system can be locked in commensurate phases at relatively high order.
In the opposite case, the scenario changes dramatically and becomes more complex.  
If $k$ is small
the mode-locking plateaus shown up in the classical model are flattened
totally by increasing the quantum effect. This result agrees
with that of a continuum limit. However, if
$k$ is relatively large, the mode-locking plateaus at rationals can be flattened, 
persistent or even enhanced by the quantum effect.
The occurrence of the incommensurate phase mode-locking depends on the
degree of irrational of $\rho$. The
mode-locking of incommensurate phase found in this paper provides us a possible way to verify the phase transition by breaking
of analyticity experimentally. This phenomenon has not yet 
been confirmed in laboratory, although it has been predicted 
theoretically for two decades. We 
hope that our findings in this paper may draw attentions from  
experimentalists.

\stars
We would like to thank Profs. S. Aubry, Gang Hu and Ding Yu Xing for
useful and stimulating discussions. This work was supported in part by
grants from Hong Kong Research Grants Council (RGC) and the Hong Kong
Baptist University Faculty Research Grant (FRG). B. Li was also supported by the Academic Research Fund of
National University of Singapore and 
H. Zhao by the National Basic Research Project ``Nonlinear Science",
China. 

%%%%%%%%%%%%%%%%%%%%%%%%%%%%%%%%%%%%%%%%%%%%%%%%%%%
\begin{figure}
%\epsfxsize=12cm
%\epsfbox{fig1.eps}
\caption{
The quantum zero point energy $f_0$ as a function of $\rho$.
Three different curves correspond to $k=0.9, 1.1 $ and $1.3$, 
respectively. The peaks happen exactly at rational numbers, while the 
minimums locate at the irrationals. The number of peaks decreases as 
$k$ is increased.} 
\end{figure}
%%%%%%%%%%%%%%%%%%%%%%%%%%%%%%

%%%%%%%%%%%%%%%%%%%%%%%%%%%%%%%%%%%%%%%%%%%%%%%%%%%
\begin{figure}
%\epsfxsize=12cm
%\epsfbox{fig2.eps}
\caption{ 
$\rho$  versus $\mu $. The $\rho$ given in this figure corresponds to that value at which $f$ takes minimum. The curves are ordered from right to left on the left hand side
of point $P$ while from right to left on the right hand side of
the point as $\tilde{\hbar }$ is increased from $0,0.05,0.1,0.2,0.3,0.4$.
} 
\end{figure}
%%%%%%%%%%%%%%%%%%%%%%%%%%%%%%%%%%%%%%

%%%%%%%%%%%%%%%%%%%%%%%%%%%%%%%%%%%%%%%%%%%%%%%%%%%
\begin{figure}
%\epsfxsize=12cm
%\epsfbox{fig3.eps}
\caption{
Specific heat $C_v$ versus $k$ for an incommensurate phase with
the golden mean $\rho$. The temperature is fixed at $T=0.2$ and $\rho =233/610$.
The curve is obviously divided into two parts. At $k<k_c=0.9716...$ 
region, $C_v$ is marginally independent of $k$, 
while at $k>k_c$ the $C_v$ decreases exponentially.
} 
\end{figure}
%%%%%%%%%%%%%%%%%%%%%%%%%%%%%%%%%%%%%%


\begin{thebibliography}{99}
\bibitem{note}
The usual definition for the golden mean is 
$[11111...]$, but
for the FK model it is identical with $[21111...]$ since the symmetry of the
system, see ref. \cite{Peyrard}, so we refer to $[21111...]$ as the golden
mean and $[31111...]$ as the silver mean and so on in the study of the FK
model.
\end{thebibliography}
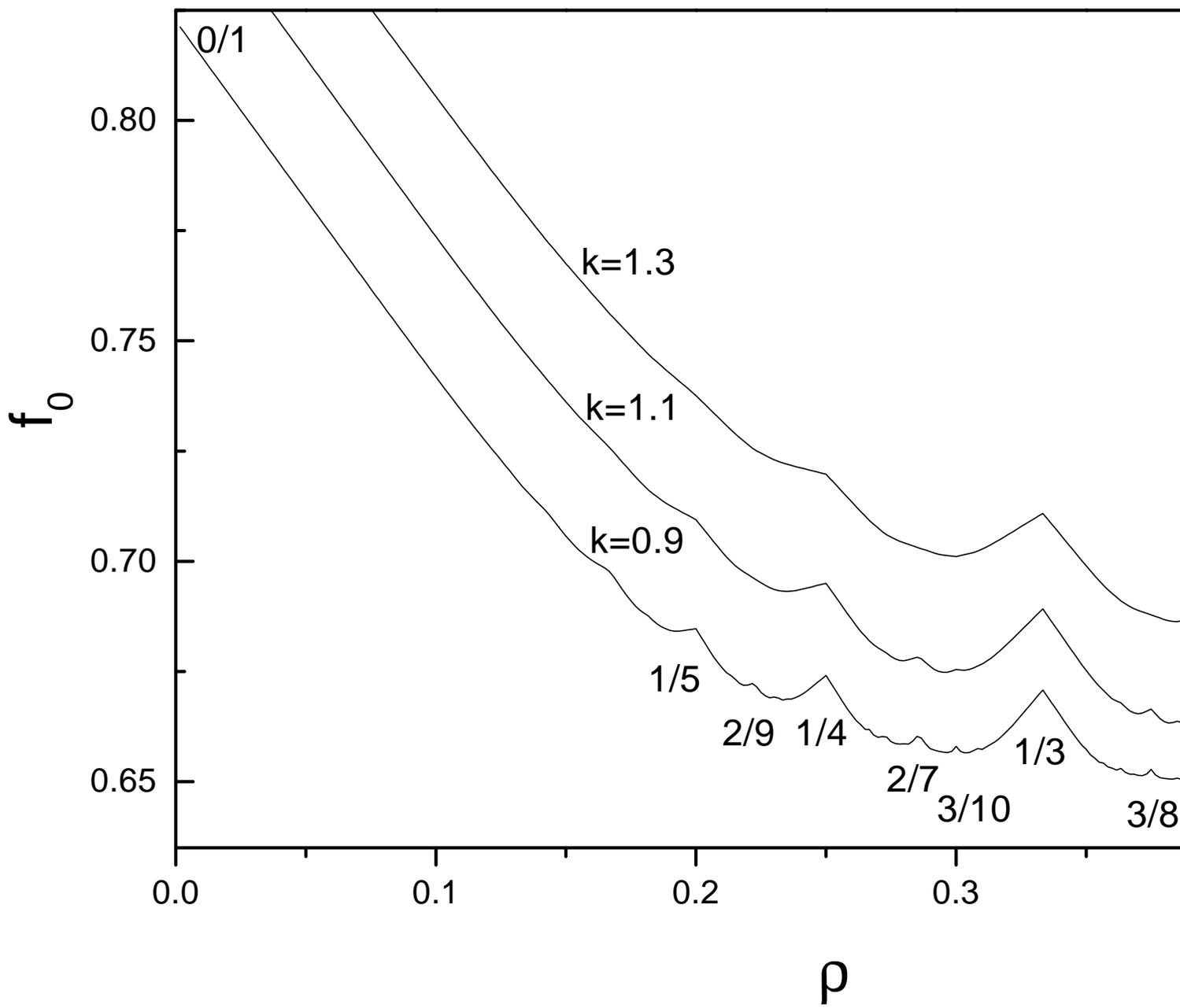
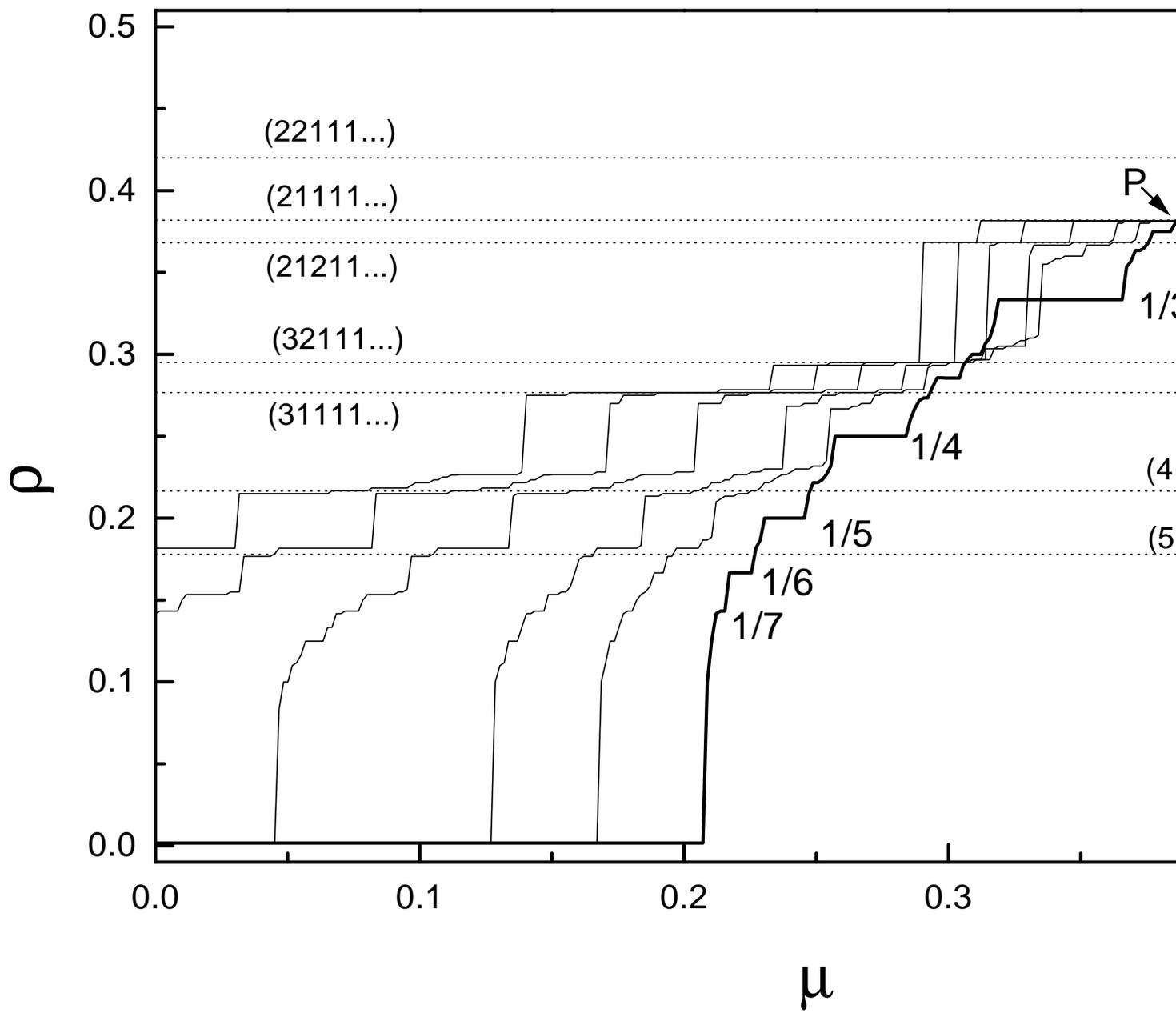
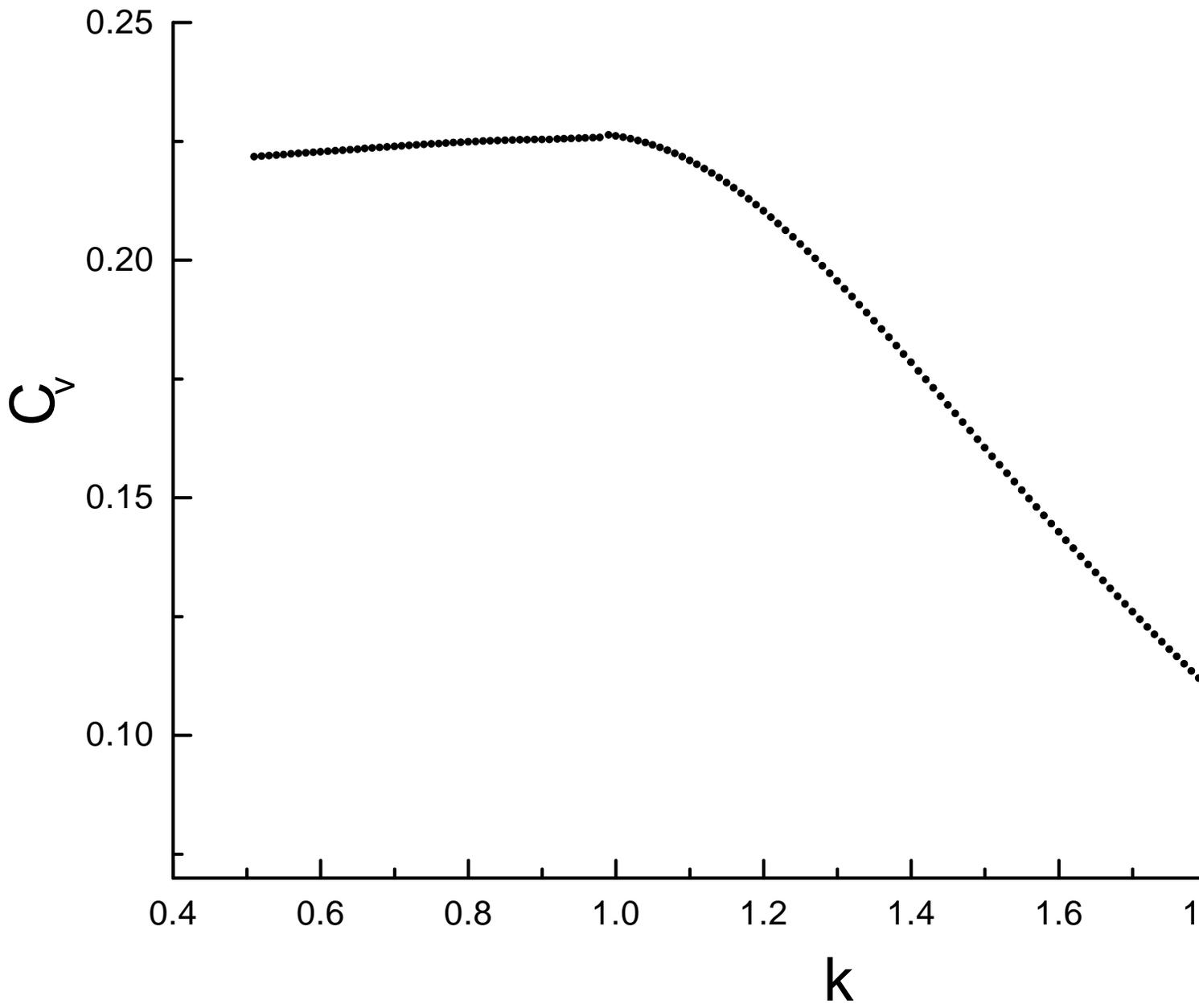
\end{document}